\documentclass{article}
\usepackage[colorlinks=true,allcolors=blue]{hyperref} \usepackage[numbers,sort&compress]{natbib}
\usepackage{flushend}  
\usepackage{lipsum} 
\usepackage{graphicx,physics}
\usepackage{geometry}
\usepackage{amsmath}
\usepackage{amssymb}
\usepackage{color}
\usepackage{booktabs}
\usepackage{xcolor}
\usepackage{colortbl}
\usepackage{tabularx}
\usepackage{stix}
\usepackage{microtype}
\usepackage[normalem]{ulem}
\usepackage[resetlabels,labeled]{multibib}
\newcites{SM}{SM References}

\definecolor{mytranscolor}{cmyk}{0.0,0.71,0.56,0.09}
\newcolumntype{a}{>{\columncolor{mytranscolor!25}}c}
\newcolumntype{B}{>{\columncolor{mytranscolor!10}}c}
\arrayrulecolor{mytranscolor!40}

\DeclareRobustCommand{\kx}{k_x}
\DeclareRobustCommand{\epl}{e^{i \kx}}
\DeclareRobustCommand{\emn}{e^{-i \kx}}

\DeclareGraphicsExtensions{.png,.jpg,.eps}

\DeclareRobustCommand{\HfigOne}{\hspace{-0.1em}\raisebox{-0.25\height}{\includegraphics[width=0.42cm]{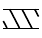}}\hspace{-0.1em}}
\DeclareRobustCommand{\HfigTwo}{\hspace{-0.1em}\raisebox{-0.25\height}{\includegraphics[width=0.42cm]{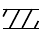}}\hspace{-0.1em}}
\DeclareRobustCommand{\HfigThree}{\hspace{-0.1em}\raisebox{-0.25\height}{\includegraphics[width=0.42cm]{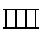}}\hspace{-0.1em}}

\newenvironment{acknowledgement}{\paragraph{Acknowledgement}}{}
\newenvironment{funding}{\paragraph{Funding}}{}
\newenvironment{authorcontributions}{\paragraph{Author Contributions}}{}
\newenvironment{conflictofinterest}{\paragraph{Conflict of Interest}}{}
\newenvironment{ethicalapproval}{\paragraph{Ethical Approval}}{}
\newenvironment{dataavailabilitystatement}{\paragraph{Data Availability Statement}}{}

\begin{document}
\restoregeometry 
\newgeometry{left=2cm,right=2cm,top=1.5cm,bottom=1.5cm}

\date{\today}

\author{
  Valerii Kachin\\
  \normalsize Institute of Informatics, Faculty of Mathematics, Informatics and Mechanics, University of Warsaw\\ \normalsize Banacha 2, 02–097 Warsaw, Poland
  \and
  Juan Camilo L\'opez Carre\~no\\
  \normalsize Institute of Informatics, Faculty of Mathematics, Informatics and Mechanics, University of Warsaw\\ \normalsize Banacha 2, 02–097 Warsaw, Poland\\
  \normalsize Institute of Theoretical Physics, Faculty of Physics, University of Warsaw\\ \normalsize Pasteura 5, 02–093 Warsaw, Poland
  \and
  Magdalena Stobi\'nska\thanks{magdalena.stobinska@gmail.com}\\
  \normalsize Institute of Informatics, Faculty of Mathematics, Informatics and Mechanics, University of Warsaw\\ \normalsize Banacha 2, 02–097 Warsaw, Poland
}

\title{Ultra-robust topologically protected edge states in quasi-1D systems}
\maketitle
\begin{abstract}
In recent years, the study of topologically non-trivial structures in one-dimensional models has been dominated by the Su--Schrieffer--Heeger model due to its simplicity in design and its ability to support edge states with robustness to disorder in couplings, protected by chiral and inversion symmetry. Here, we present a novel study on a zigzag quasi-one-dimensional model, which supports topologically protected edge states without relying on conventional symmetries. Our model utilises next-neighbour couplings to mediate edge states and is simultaneously resilient to dissipation, couplings and on-site energy disorders. In order to understand the topological properties of this model, we introduce a novel way to demonstrate the bulk-boundary correspondence of the edge states and construct a topological invariant that returns quantized values. Our study sheds light on the possibility of constructing topological phases in new ways, even in the absence of conventional symmetries, and opens up new avenues for research in this field. In addition, we demonstrate a possible photonic realization of these models with the help of an orbital-induced synthetic flux.
\end{abstract}

\restoregeometry 
\newgeometry{twocolumn, margin=2cm}

\section{Introduction} 

Topologically protected one-dimensional edge states are quantum states of matter whose wave function is concentrated at the edges of a 1D lattice, and vanishes in its bulk. Lattice symmetries shield them from noise, crystal defects, or chemical composition changes~\cite{hasan2010colloquium,qi2011topological}. Their energy levels lie inside the energy gap between the valence and conduction bands. Beyond fundamental studies~\cite{vonKlitzing2020,RevModPhys.95.011002,RevModPhys.87.137,HasanNatureReview2021,NennoAxion2020}, they open prospects for fault-tolerant quantum computation~\cite{KITAEV20032, RevModPhys.80.1083, pachos_2012, PhysRevB.100.045414, SciPostPhys.3.3.021}, robust information processing~\cite{Mittal:16, Rechtsman:16, 10.1063/5.0050672,Alicea2011, Blanco-Redondo2018, flamini2018photonic} and building numerous devices~\cite{ Jean-LucTambasco2018, Yue2019adv, TopologicalValleyPhotonics, Price_2022, PhysRevApplied.20.024076}.

The best known edge states are protected by the chiral symmetry, described by the Su--Schrieffer--Heeger (SSH) chain model~\cite{PhysRevLett.42.1698,PhysRevB.21.2388,PhysRevLett.89.077002}. Here, an electron can either hop between the two sites within the unit cell, or to an adjoining one, with strengths~$v$ and~$w$, respectively. If $w>v$, topologically protected states appear. 

\begin{figure}[t]
 \centering
 \raisebox{1.3cm}{(a)}
 \includegraphics[width=7.5cm]{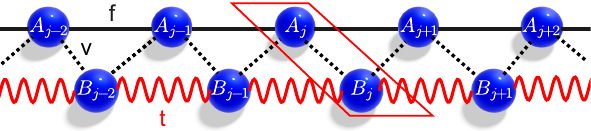}\\
 \raisebox{1.3cm}{(b)}
 \includegraphics[width=7.5cm]{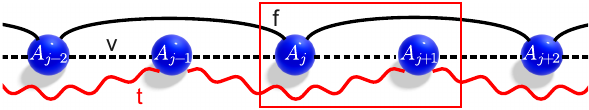}
 \caption{(a) Quasi-1D zigzag structure with interacting chains $A$ and $B$ connected by strength $v > 0$ (dashed line). The coupling strengths between the sites within chain $A$ is $f > 0$ (black line), while within $B$ is $t < 0$ (red line). (b) Our system can be represented as a chain with next-neighbor and long-range interactions. An elementary two-site unit cell is shown as a red frame.}
 \label{fig:fig1}
\end{figure}

Various models supporting edge states have been proposed, including SSH generalizations~\cite{PhysRevB.89.085111, Maffei_2018,RYU2022114941,LEE202296, PhysRevB.99.035146}, quasicrystal structures~\cite{PhysRevLett.110.076403, kraus2016quasiperiodicity, Zilberberg:21}, the Aubry--Andr\'e--Harper (AAH) model~\cite{aubry1980analyticity, PhysRevLett.109.106402}, and Anderson topological insulators~\cite{PhysRevLett.102.136806, stutzer2018photonic}. They exhibit robustness against disorders in couplings that result from variations in distances between quantum particles~\cite{TopologicalPhotonicsonaSmallScale, 2021SegevBandres}. However, while the previous research addressed spectral disorders and on-site imperfections, particularly in non-dissipative systems~\cite{PhysRevB.103.024106, PhysRevResearch.3.033012}, the effects of site-dependent losses and phase disorder have received scant attention. 
These imperfections can significantly degrade edge state localization, induce energy gap closure, and prompt a shift from topological to conventional matter~\cite{hasan2010colloquium,qi2011topological}. Therefore, developing edge states resilient to these defects is crucial.

This letter presents the emergence of remarkably resilient topologically protected edge states in a quasi-1D zigzag structure where some lattice couplings are negative (with phase $\pi$), even \emph{in the absence} of chiral symmetry. These states maintain their nature despite simultaneous disorders in lattice couplings, variations in on-site phases and potentials, and dissipation. Negative couplings play a crucial role in their stabilization, localization, and appearance of energy gaps. We introduce a novel approach that allows us to restore the bulk-boundary correspondence, and to compute a regularized topological invariant. It establishes a continuous mapping between the original model and the ``parental model'', with restored chiral symmetry. To this end, we construct in the Fourier space a special variant of the unit cell that we call ``weighted''. 

Negative couplings have gained attention for achieving non-trivial topological phases and creating energy gaps~\cite{BenalcazarQEMI2017,Fu2020,KremerNatC2020,Ni2020,PhysRevB.105.184108,PhysRevApplied.16.024032,Jiang:23,Zhenzhen:23}. An example is the quadrupole topological insulator (QTI) exhibiting corner states limited by robustness to interaction strength disorders~\cite{BenalcazarQEMI2017,Peterson2018}. Negative couplings naturally emerge in quantum particle interactions in alternative media: atoms in optical potentials~\cite{RevModPhys.83.1523,Li2013}, arrays of microwave resonators~\cite{PhysRevLett.110.033902,Peterson2018}, optical waveguide lattices~\cite{Nam:10, Zeuner:12, Zhenzhen:23, PhysRevLett.116.213901, Pocock2018, Fu2020, PhysRevA.102.023505,Schulz2022,Jiang:23}, acoustic cavities~\cite{Ni2020, PhysRevLett.124.206601, PhysRevB.105.184108}, synthetic dimensions~\cite{Dutt2020,Ni::Synth2021}, or resonant electric circuits~\cite{Bao2019,Liu-Ma-Zhang2020}. We present an approach for photonic realization of proposed quasi-1D model with construction of negative couplings between waveguides through synthetic orbital-induced flux, which have been used before mostly in formation of nearest-neighbor interactions. 

Our results pave the way for realistic implementation of topological quantum computing, information processing, and quantum transport on numerous platforms, including photonic, acoustic, mechanical metamaterials, and resonant electric circuits.

\section{Edge states}
Let us consider the zigzag structure shown in Fig.~\ref{fig:fig1}a where one of the interaction strength is negative. It is composed of two interacting topologically trivial chains $A$ and $B$, and it is described by a finite tight-binding Hamiltonian
\begin{equation}
 \label{eq:Hamiltonian}
H\!=\!\!\! \sum_{j=1}^{N-1} \bigl(va_j^\dagger b_{j+1} + fa^\dagger_j a_{j+1} + t b^\dagger _j b_{j+1}\bigr)+\!\!\sum_{j=1}^{N} v a^\dagger_j b_j +h.c.,
\end{equation}
where $N$ is the number of sites on each chain, $v \ge 0$ is the hopping amplitude between them, $f \ge 0$ ($t\le 0$) is the hopping amplitude within the chain $A$ ($B$); and $a_j$ ($a_j ^\dagger$) and $b_j$ ($b_j ^\dagger$) denote annihilation (creation) operators at the $j$-th site of the chain $A$~($B$). Note that we did not made any assumption of the statistics that the operators~$a$ and $b$ follow.

\begin{figure}[t]
 \centering
 \includegraphics[scale=0.59]{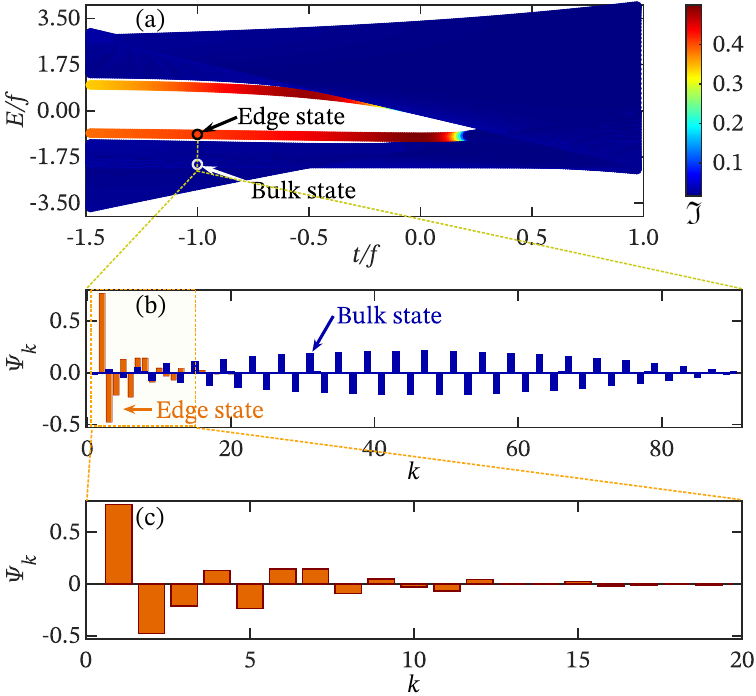}
 \caption{(a) Energy band structure of the zigzag system with $N=90$ sites computed for $v/f=1$, as a function of the interaction strengths' ratio $t/f$. The edge (red) and bulk (blue) states here and further are highlighted using the inverse participation ratio (IPR) $\mathfrak{I}$. (b) Mode amplitude $\Psi_k$ on site $k$ of the edge and bulk states' wave function evaluated for $t/f=-1$. (c) Profile of an edge state eigenmode shown for the first $20$ sites. }
 \label{fig:fig2}
\end{figure}

Our system can be viewed as a chain with uniform strength of next-neighbor (nn) interactions, and two opposite in phase strengths of next-next-neighbor (nnn) interactions, Fig.~\ref{fig:fig1}b. The presence of long-range repulsive and attractive interactions renders this model distinct from the SSH and AAH. All its non-trivial topological properties are due to the nnn-interactions; a model without them becomes a topologically trivial gapless chain. The analysis of energy bands of $H/f$ in terms of the inverse participation ratio $\mathfrak{I}$ (IPR)~\cite{THOULESS197493} (see Supplemental Material, SM, Section S1) for a chain of $90$ sites shown in Fig.~\ref{fig:fig2}a reveals the presence of two bands if $v^2 > t f(t+1)^2(f+1)^2$ and $f\neq t$, and edge states within the gap if $v^2/4 > tf$ and $v\neq0$. The distance between the edge state and the closest band is larger, and the above inequalities always hold true simultaneously if $t/f < 0$. Figs.~\ref{fig:fig2}b--c show the structure of edge states for $t=-f$, when the system acquires chiral symmetry.

To describe the bulk properties of the two-band spectrum shown in Fig.~\ref{fig:fig2}, it is sufficient to consider a unit cell that consists of two sites, $A$ and $B$, and the corresponding bulk Hamiltonian in the continuum space
\begin{equation}
 H_2(\kx) = \begin{pmatrix}
 	2f \cos{\kx}& v\bigl(1+\emn\bigr)\\
 v\bigl(1+\epl\bigr)& 2t\cos{\kx}
 \end{pmatrix}.
 \label{eq:eq2}
\end{equation}
Then, the bulk spectrum equals  $\varepsilon= (f + t) \cos \kx \break \pm \sqrt {(f - t)^2 \cos^2 \kx + 2 v^2 (1 + \cos \kx)}$. If $t=-f$, it exhibits the chiral symmetry, $\Gamma_2 H_2(\kx) \Gamma_2^*=-H_2(\kx)$, where $\Gamma_2=\sigma_y T_2$, and the winding number is $W=0$~\cite{ryu2010topological}. In a general case, however, the $2\times2$ bulk model in Eq.~(\ref{eq:eq2}) possesses neither chiral nor inversion symmetry. Thus, both the Zak phase and the winding number are not quantized, and the use of bulk-edge correspondence is not justified.

\section{Topological invariant}
\begin{figure} 
    \centering
    \includegraphics[width=8cm]{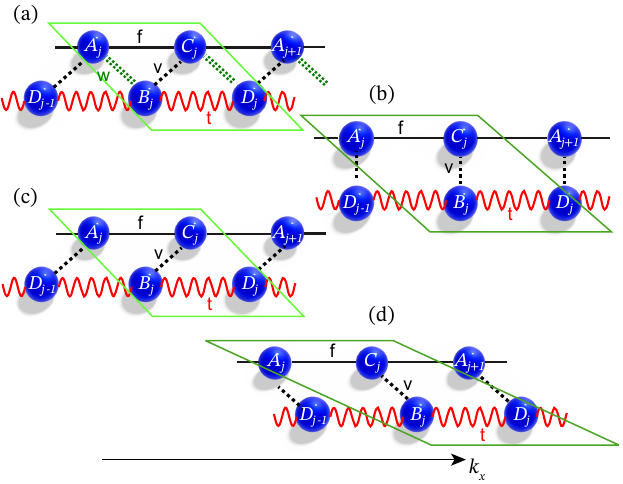}\\
    \caption{``Parental'' model for the quasi-1D zigzag structure. (a) We generalize the zigzag model by allowing the chains $A$ and $B$ to interact with two alternating strengths, $v$ and $w$. For $f=t=0$ it restores the SSH model. (b) If $w=0$, the model amounts to a ladder with four-site non-trivial unit cell. (c) and (d) are two geometrical interpretations of the quasi-1D ladder structure. Unit cells at (b), (c) and (d) correspond to Hamiltonians $H_4\left(\HfigThree\right)$, $H_4\left(\HfigTwo\right)$ and $H_4\left(\HfigOne\right)$, respectively.}
    \label{fig:fig3}
\end{figure}

We propose to construct a ``parental'' model that features the chiral symmetry, connected to our original zigzag model by a continuous transform, Fig.~\ref{fig:fig3}a, with a four-sites unit cell. Here, this ``parental'' model is a ladder system, Fig.~\ref{fig:fig3}b, with initial setting $w=0$, coinciding with Fig.~\ref{fig:fig1}a for $w=v$. Since transition occurs without closing the energy gap, the two models can be continuously connected by a mapping $f(H_4):=(\Gamma_4 H_4 \Gamma_4^*-H_4)/2$, where $\Gamma_4=\sigma_z\otimes I_2$ is the chiral symmetry operator for the ``parental'' model.

Let $H_4^{\text{par}}$ and $H_4^\text{zig}$ are the bulk Hamiltonians of the ``parental'' and the zigzag models, respectively, both defined on a four-site unit cell. Then, $f(H_4^\text{par})=f(H_4^\text{zig})=H_4^\text{par}$. This allows us to draw conclusions about topological edge states in the zigzag structure solely based on the analysis of the ladder system. The construction of the winding number for the latter requires a regularization step: the quasi-1D nature necessitates a correct geometrical interpretation of couplings $v$ (perpendicular to the selected momentum direction) between two chains.

Let us now denote the unit cells of the ladder systems from Fig.~\ref{fig:fig3}b--d as \HfigThree, \HfigTwo, and \HfigOne, respectively. While Hamiltonian $H_4\left(\HfigThree\right)$ requires definition, the winding number integral calculated for Hamiltonians $H_4\left(\HfigTwo\right)$ and $H_4\left(\HfigOne\right)$ yields $W=1$ in the presence of in-gap edge states, and does not converge in their absence. Thus, we construct a ``weighted'' unit cell, and the corresponding Hamiltonian $H_4\left(\HfigThree\right)$
\begin{equation}
    H_4\left(\HfigThree\right) = \frac{H_4\left(\HfigTwo\right) + \xi H_4\left(\HfigOne\right)}{\xi+1},
\end{equation}
which properties in the basis of $\Gamma_4$ are determined by the upper-right block
\begin{equation}
h_2(\kx, \xi)=\begin{pmatrix}
f\bigl(1+\emn\bigr)& v \frac{\emn +\xi \epl}{\xi+1} \\
v & t\bigl(1+\emn\bigr)
\end{pmatrix}.
    \label{eq:eq4}
\end{equation}
It also governs the winding number which takes the form
\begin{equation}
    W=\frac{1}{i 2 \pi} \int_{BZ} d\kx \textrm{Tr} \left(h_2(\kx,\xi)^{-1}\frac{d}{d \kx} h_2(\kx,\xi)  \right)=\nonumber
\end{equation}
\begin{equation}
    =\int_0^{2\pi} \frac{d\kx}{2\pi}\frac{ \frac{2 ft}{v^2}(\xi+1) \left(1+\emn\right)+  \left(\xi e^{2 i \kx}-1\right)}{\frac{2 ft}{v^2}(\xi+1)  (1+\cos{\kx}) - \left(\xi e^{2 i \kx}+1\right)}.
    \label{eq:eq5}
\end{equation}
For $\xi \ll 1$, the integral in $W$ does not converge. However, larger values of $\xi$ lead to the identification of non-trivial region, namely $W=1$ for $t \in (-\infty;v^2/(4f))$ and $W=0$ for $t \in [v^2/(4f); \infty)$.

\begin{figure}
 \includegraphics[scale=0.59]{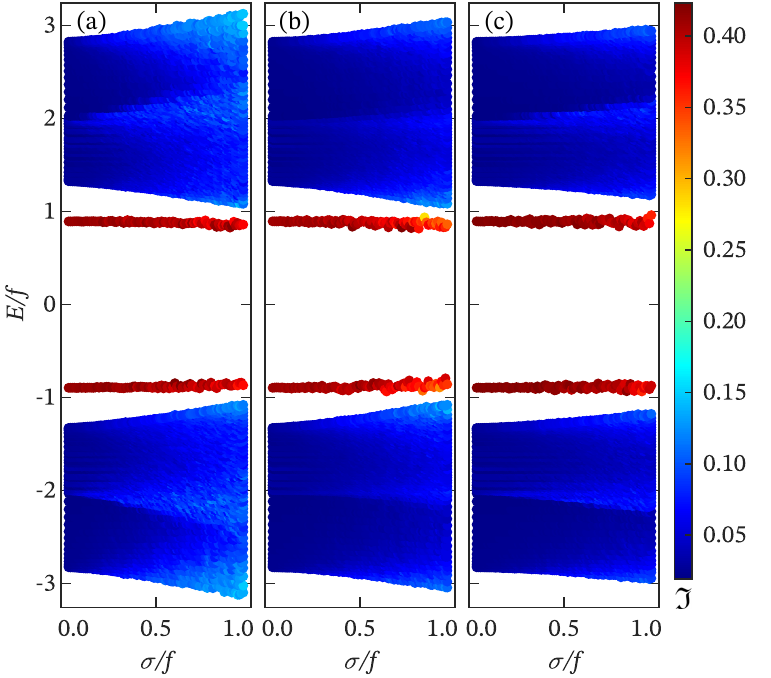}
 \caption{Energy eigenmodes computed for a chiral ($t/f=-1$) zigzag system with $N=90$ sites and $v/f = 1$, averaged over 30 random instances of a disorder in (a) nn interactions, (b) nnn interactions, and (c) on-site energy, as a function of the disorder strength $\sigma/f$. The findings stem from analyzing the finite Hamiltonian $H_{\sigma} = H + \sigma D$, featuring the tight-binding Hamiltonian $H$ and the disorder operator $D$.}
 \label{fig:fig4}
\end{figure}

\section{Robustness to disorders}
We explored the robustness of edge states in the zigzag system against disorders in the strengths of the nn and nnn interactions, and against the spectral disorder, by computing the energy spectrum of the Hamiltonian $H_\sigma = H + \sigma D$, where $H$ is defined by Eq.~(\ref{eq:Hamiltonian}), $\sigma$ is a constant that quantifies the strength of the disorder, and~$D$ is a matrix that defines the type of disorder. The matrix elements $D_{l,m}=\Delta d  \delta_{l\pm 1,m}$ and $D_{l,m}=\Delta d \delta_{l\pm 2,m}$ represent the disorder in the nn and nnn coupling constants, while $D_{l,m}=\Delta d  \delta_{l,m}$ model the spectral disorder. We sampled $\Delta d$ uniformly in range $[-0.5; 0.5]$. 

Fig.~\ref{fig:fig4} depicts the spectrum of $H_{\sigma}$ computed for the chiral ($t/f=-1$) zigzag chain that was averaged over 30 random instances of a disorder in (a) next-neighbor, (b) long-range interactions, and (c) on-site energy spectrum, for $N=90$ sites and $v/f = 1$. For each disorder type, as its strength increases, the edge states become less localized and the the bulk bands broaden. This effect becomes visible when disorders are as strong as the system interaction, $\sigma/f=1$. 

The edge states also persist in a zigzag system without the chiral symmetry. Although they reveal similar level of robustness, the positive mode edge state starts to delocalize at smaller disorders' strength, $\sigma/f=0.5$ for $t/f=-0.7$ and $\sigma/f=0.3$ for $t/f=-0.3$.  Edge states persist in chains with odd $N$, exhibiting degenerate energy levels, and higher disorders induce a minor spectral spreading of these states (see SM, S2 and S3).

\begin{figure}
 \centering
 \includegraphics[scale=0.59]{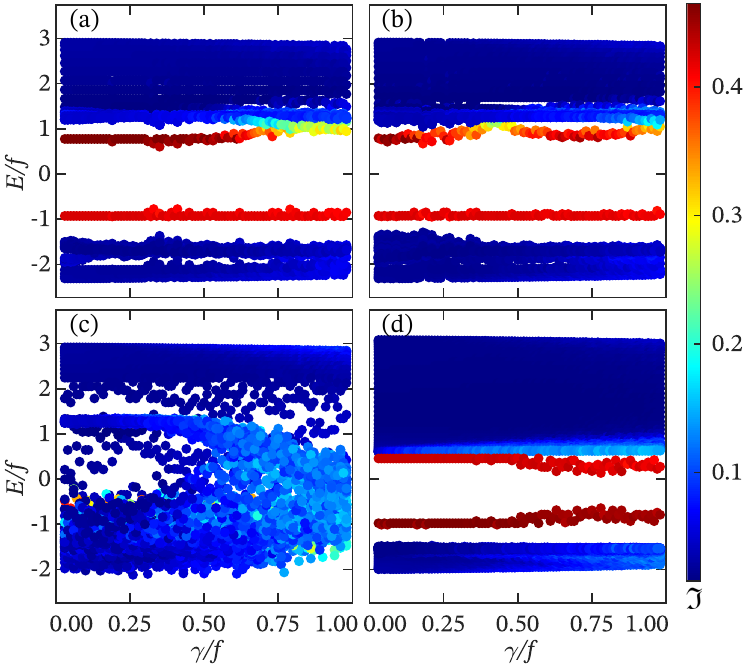}
 \caption{Eigenmodes of a finite zigzag structure, averaged over 30 random realizations of loss, as a function of loss strength $\gamma /f$ for $v/f = 1$. Panels (a), (b), and (c) are computed for $t/f=-0.7$, with the numbers of sites $N=40$, $N=60$, and $N=80$, respectively. In (d), the choice of $t/f=-0.3$ facilitates robust edge states for longer chains ($N=90$). Optimizing system parameters allows well-localized edge states even with losses’ strength as high as the system interaction. The findings stem from analyzing the finite Hamiltonian $H_{\gamma}= H + i\gamma D'$, featuring the tight-binding Hamiltonian $H$ and the loss operator $D'$.}
 \label{fig:fig5}
\end{figure}

\section{Robustness to losses}
We model dissipation in the zigzag structure by a finite non-hermitian Hamiltonian $H_\gamma = H + i\gamma D'$, where $\gamma$ represents losses and $D'$ is a diagonal matrix with elements $D'_{l,m}=\Delta d  \delta_{l,m}$. We sample $\Delta d$ uniformly from $[-0.5; 0.5]$.

\begin{figure}
    \centering
    \includegraphics[scale=0.59]{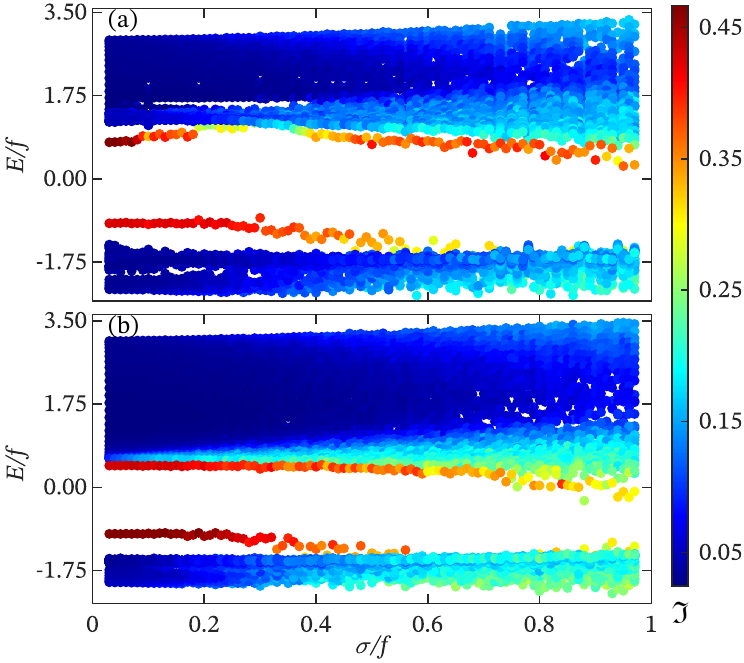}
    \caption{Our main result: robust topologically protected edge states in a zigzag system affected simultaneously by losses and disorders in next-neighbor and long-range interactions, and on-site energies; all quantified with strength $\sigma$. Remarkably, the edge states remain well localized  for the disorders and losses comparable with the system interaction strength, $\sigma/f \le 0.3$. The plot is computed for a chain of $N=60$ sites, $v/f=1$, (a) $t/f=-0.7$, (b) $t/f=-0.3$.}
    \label{fig:alltypesofdisorder}
\end{figure}

Fig.~\ref{fig:fig5} shows eigenmodes of $H_\gamma$ computed for several zigzag chains, and averaged over 30 random instances of losses, as a function of the loss strength $\gamma /f$. We observe that the stability of edge states depends on $N$ along with ratios $v/f$ and $t/f$. While larger $t/f$ increases the robustness of edge states for bigger $N$, it simultaneously reduces the width of the energy gap between the edge states and bulk bands (see SM, S4). Moreover, the losses do not break the degeneracy of the edge states for a system with odd $N$. (a), (b), and (c) highlight the system's vulnerability to losses as the number of sites increases for $t/f=-0.7$. (d) demonstrates the robustness of edge states for a chain with $t/f=-0.3$ and $N=90$. As before, $v/f=1$.

\section{Main result}
Our main result is demonstrated in Fig.~\ref{fig:alltypesofdisorder}. It shows ultra-robust topologically protected edge states obtained for the zigzag structure affected simultaneously by dissipation and all three types of disorders considered. It was obtained by an averaged simulation of the cumulative Hamiltonian $H_{\sigma} + H_\gamma$, where for the simplicity of computations, all the imperfections had the same strength $\sigma$. The computations were performed for $N=60$ sites with $v/f=1$, for $t/f=-0.7$ -- panel (a), and $t/f=-0.3$ -- panel (b). Remarkably, the edge states remain well localized for the disorders and losses comparable with the system interaction strength, $\sigma/f \le 0.5$. Here, the negative coupling is vital for securing a distinct, resilient edge state against losses and various disorders in zigzag models of different sizes. While a certain level of resilience can be attained by relying solely on positive couplings, this comes at the cost of a smaller gap and potentially weaker localization of the edge state, allowing for simultaneous excitations in both edge and bulk states (see SM, S5). Tab.~\ref{tab:1} compares the robustness of our edge states to the SSH, AAH, and QTI models.

\begin{table}
\begin{tabular}{@{}cBBBa@{}}\toprule
& SSH & AAH & QTI & Our model\\ \midrule
\rowcolor{mytranscolor!0}
\multicolumn{5}{c}{System properties} \\ \cmidrule{1-5}
Couplings & positive & positive & positive & positive \\[-0.5ex] 
&&& \& negative & \& negative\\[0.5ex] \cmidrule[0.005em](lr){1-1} 
Topological & 1-2 & 1-4 & 1-4 & 1-2 \\[-0.5ex] 
states & at $\varepsilon=0$ 
& at $\varepsilon\neq 0$ 
& at $\varepsilon=0$ 
& at $\varepsilon\neq 0$
\\
[0.5ex]\cmidrule[0.005em](lr){1-1}
Self-energy & -- & yes & -- & yes\\ [0.5ex]\cmidrule{1-5}
\rowcolor{mytranscolor!0}
\multicolumn{5}{c}{Robustness to disorder type} \\ \cmidrule{1-5}
In couplings & \checkmark & \checkmark & \checkmark & \checkmark \\ [0.5ex]\cmidrule[0.005em](lr){1-1}
On-site energy & -- & \checkmark & -- & \checkmark\\ [0.5ex]\cmidrule[0.005em](lr){1-1}
Losses & -- & -- & -- & \checkmark\\[-0.5ex]
(dissipation)&&&& \\
\bottomrule
\end{tabular}
\caption{Robustness of topologically protected edge states obtained in the SSH, AAH, and quadrupole topological insulator (QTI) models, with these observed in the discussed zigzag structure with negative couplings. $\varepsilon$ denotes the energy level of an edge state.}
\label{tab:1}
\end{table}

\section{Implementation}

\begin{figure}
    \centering
    \includegraphics[width=8cm]{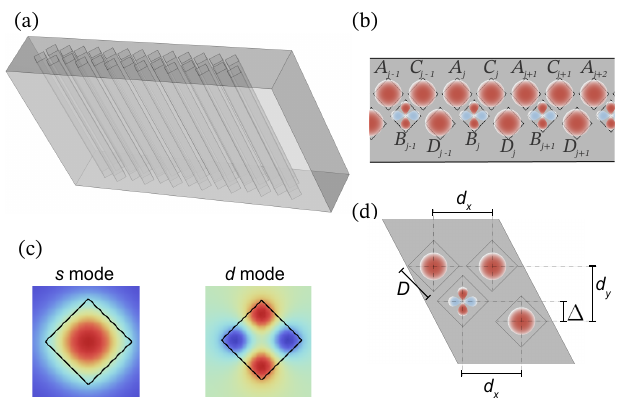}
    \caption{The photonic waveguide structure designed with orbital-induced synthetic flux is depicted as follows: (a) a 3D schematic structure of the zigzag model and (b) a cross-sectional view. Figure (c) showcases the fundamental mode pattern $E_x$ (at $\lambda= 1.55$ $\mu m$) of the individual square waveguides with refractive indexes $n_s=1.542$ and $n_d=1.5468$ and refractive index of the cladding $n_{cl}=1.54$. Figure (d) provides an illustration of the unit cell and parameters utilized to control coupling strengths. The side length of the isolated square waveguide is $D= 12.4$ $\mu m$. The distances between the centers of waveguides are $d_x=20$ $\mu m$ and $d_y=16$ $\mu m$, with a shift parameter adjusted to manipulate diagonal couplings, denoted as $\Delta=4$ $\mu m$.}
    \label{fig:setupscheme}
\end{figure}

\begin{figure}[ht]
    \centering
    \includegraphics[width=8cm]{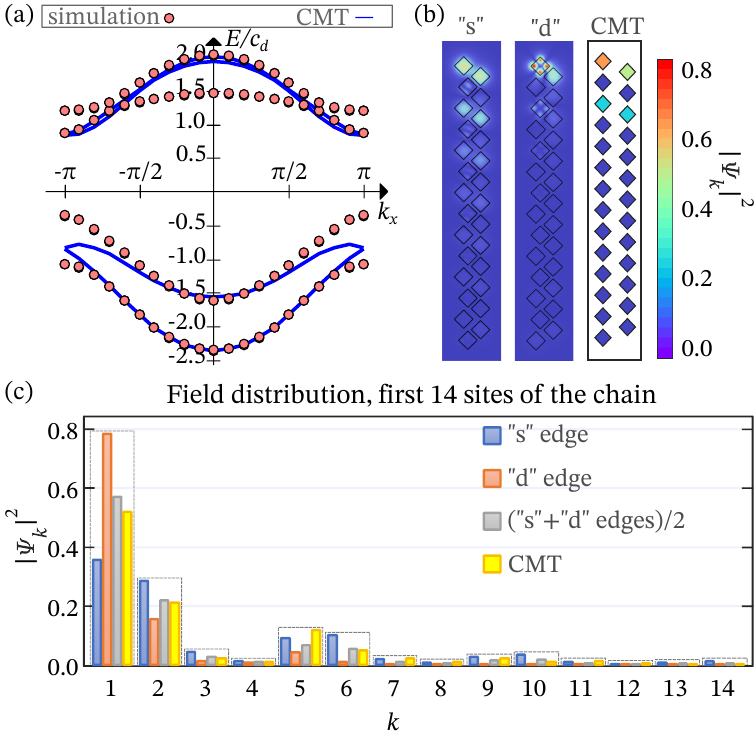}
    \caption{Spectrum properties of the photonic waveguide structure designed with orbital-induced synthetic flux: (a) The band structure of the zigzag model is shown. The blue solid lines depict results obtained from simulations, while the red circles represent results from CMT. In Figure (b), unit-norm field distributions (Ex) of edge modes marked by "s" and "d", corresponding to different types of edges, along with the edge mode amplitude distribution of the CMT model with $v/f=0.75$, $t/f=-1.27$. In Figure (c), a histogram displays the profiles of a unit-norm edge state eigenmode for the first 14 sites. Blue and red indicate the profiles of 's' and 'd' edges of the photonic structure, while grey represents their average value, which closely aligns with the CMT results shown in yellow.}
    \label{fig:setup_profiles}
\end{figure}

In the preceding sections, we demonstrated the robustness and topological properties of the edge states within the zigzag model. In this section, we proceed to showcase the implementation and potential realization of these theoretically predicted phenomena in 3D photonic waveguide arrays through orbital-induced synthetic flux  \cite{Zeuner:12, Jiang:23, Zhenzhen:23}. The geometry of the waveguide structure, as depicted in Fig.\ref{fig:setupscheme}, consists of fourfold symmetric waveguides with square cross-sections, supporting both $s$ and $d$ modes. A cross-sectional slice is illustrated in (b). Notably, the parameters of the waveguides are as follows: a side length of $D= 12.4 \mu m$, refractive indexes of waveguides $n_s=1.542$ and $n_d=1.5468$, and a refractive index of cladding $n_{cl}=1.54$, chosen to ensure that individual $s$ and $d$ modes possess the same propagation constant $\beta_0=6.2464 \mu m^{-1}$. The coupling strength and polarity are determined by the overlap of the decoupled base states and can be adjusted by spatial parameters such as $dx$, $dy$, and $\Delta$. Fig.\ref{fig:setupscheme}(c) presents the fundamental modal electric field $E_x$ (at $\lambda =1.55 \mu m$) of the decoupled square waveguides, represented by $s$ and $d$ orbitals, respectively. These results were obtained from numerical simulations using the finite element method (FEM) in COMSOL Multiphysics\texttrademark{} 6.1. Fig.~\ref{fig:setupscheme}(d) illustrates the unit cell of the waveguide lattice, consisting of four square waveguides; negative coupling arises from the interaction between $s$ and $d$ modes. Diagonal site couplings have been spatially adjusted to achieve relatively similar strength by $\Delta$ shift. The presented design supports band gap and edge states that demonstrate behavior similar to the proposed theoretical model, which could be seen from Fig.~\ref{fig:setup_profiles}.


The couple mode theory (CMT) analyses, yielding coupling values $v/f=0.75$ and $t/f=-1.27$, align well with the full-wave simulations conducted using FEM. This agreement is evident when comparing the bulk bands in Fig.\ref{fig:setup_profiles}(a). The same edge state in the photonic waveguide structure can be formed with both $s$ and $d$ modes, exhibiting similar localization characteristics consistent with CMT predictions, as depicted in Fig.\ref{fig:setup_profiles}(b). A detailed comparison of edge mode profiles on the first 14 sites of zigzag structure in Fig.~\ref{fig:setup_profiles}(c) demonstrates that the average values of "s" and "d" edge states closely match those predicted by CMT.

Furthermore, the same scheme of orbital-induced synthetic flux can be realized with circular waveguides or by achieving effective negative coupling through the insertion of an auxiliary defect waveguide with a precisely tuned refractive index between the two original waveguides \cite{Zeuner:12, Jiang:23}. Consequently, coupling noise can be controlled by adjusting waveguide spacing, on-site noise can be modulated by varying waveguide widths, and system loss can be adjusted by altering waveguide etch depth.

Alternatively, other approaches can be considered, such as utilizing acoustic metacrystals \cite{Ni2020} to fabricate the proposed chain through 3D printing, where negative couplings are formed by a special choice of unit cell connections. Additionally, superconducting circuits \cite{PhysRevB.93.054116} can be employed, where adjacent sites in the lattice are connected through SQUIDs and magnetic flux can assist in manipulating the coupling sign. Resonant electric circuits offer another avenue, implementing different couplings with different sets of capacitors and inductors.

\section{Discussion and conclusions}
We demonstrated an extreme robustness of topologically protected edge states emerging from the interacting quasi-1D zigzag structure, where some of the couplings are negative. These states remain well localized in the simultaneous presence of dissipation and disorders in short- and long-range interactions, as well as in on-site energies, whose strengths are comparable with interactions in the system. We proposed photonic designed and also identified several physical platforms that are capable of testing our results.

We proved non-trivial topological nature of the edge states using the bulk-boundary correspondence, despite the lack of chiral symmetry in the system. To this end, we proposed a new method that relates the original zigzag system via a continuous mapping to a ladder model featuring  chiral symmetry, which is investigated in a larger, specially constructed unit cell.

The model's simplicity and generality can make it a cornerstone for future practical quantum topology-based applications. Realization of negative couplings plays a pivotal role in broadening the range of system parameters, enabling it to support edge states, develop new symmetries, and control the energy band gap size.

\begin{acknowledgement}
V.K. thanks Alexander Khanikaev for the initial discussions and Diana Shakirova for COMSOL Multiphysics \texttrademark{} advices.
\end{acknowledgement}

\begin{funding}
V.K. and M.S. were supported by the European Union’s Horizon 2020 research and innovation programme under the Marie Sk\l{}odowska-Curie project ``AppQInfo'' No.~956071. M.S. was supported by the National Science Centre ``Sonata Bis'' project No.~2019/34/E/ST2/00273, and the QuantERA II Programme that received funding from the European Union’s Horizon 2020 research and innovation programme under Grant Agreement No.~101017733, project ``PhoMemtor'' 
\newline No. 2021/03/Y/ST2/00177. 
\end{funding}

\begin{authorcontributions}

V.K. conceived the original concept of the work, designed and studied the structures, developed the described techniques and performed full-wave simulation. J.C.L.C. assisted with discussions. M.S. supervised the project. V.K. and M.S. wrote the manuscript. All authors have discussed the results, have accepted responsibility for the entire content of this manuscript and approved its submission. 
\end{authorcontributions}

\begin{conflictofinterest}
 \emph{Authors state no conflict of interest.}
\end{conflictofinterest}

\begin{ethicalapproval}
\emph{The conducted research is not related to either human or animals use.}
\end{ethicalapproval}

\begin{dataavailabilitystatement}
\emph{The datasets generated and/or analyzed during the current study are available from the corresponding author upon reasonable request.}

\end{dataavailabilitystatement}

\bibliographystyle{unsrt}
\bibliography{our}

\restoregeometry 
\newgeometry{left=2cm,right=2cm,top=1.5cm,bottom=1.5cm}

\setcounter{figure}{0}

\renewcommand{\thefigure}{SM\_\arabic{figure}}
\renewcommand{\thesection}{SM\arabic{section}}

\begin{center}
  {\LARGE\bfseries Supplemental Material: Ultra-robust topologically protected edge states in quasi-1D systems}
\end{center}
\setcounter{section}{0}
\section{Inverse participation ratio}

\noindent
To quantify mode localization properties, we calculate an inverse participation ratio $\mathfrak{I}$ (IPR), defined in the following way~\citeSM{THOULESS197493_1}
\begin{equation}
    \mathfrak{I} =\left(\sum\limits_{m}\,|\psi_{m}|^4\right)\,\left(\sum\limits_{m}\,|\psi_{m}|^2\right)^{-2}.
\end{equation}
Here, $\psi_{m}$ represents a mode amplitude on site $m$ in the 1D system. The mode corresponds to e.g., a field strength, voltage, pressure, or other relevant physical quantities characteristic to a physical implementation under consideration. The summation is performed over all $N$ sites within the chain.

\section{Robustness for chains with even number of sites in the non-chiral cases}

\begin{figure}[h!]
    \includegraphics[scale=0.56]{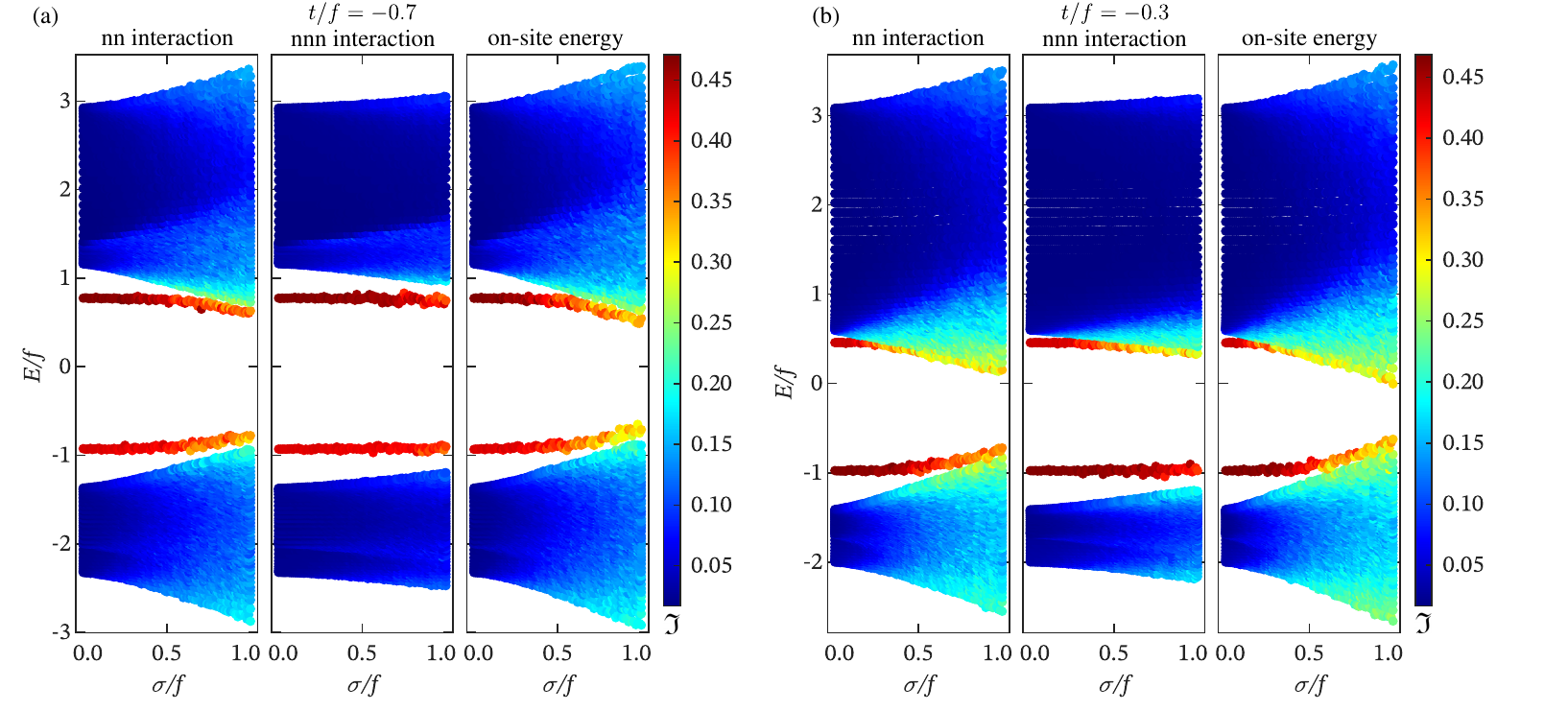}
    \caption{Energy eigenmodes computed for the zigzag system with $N=90$ sites, while setting~$v/f=1$ and (a)~$t/f=-0.7$  and~(b)~$t/f=-0.3$. Each case was averaged over 30 random instances of a disorder $\sigma$,  for (left to right) the next-neighbour (nn) interactions, long-range (nnn) interactions, and the on-site energy.}
    \label{fig:fig1sm}
\end{figure}

\noindent 
In the figures presented in the main text, we illustrate the robustness properties of a chain consisting of an even number of sites in the chiral case. Interestingly, nonchiral cases exhibit nearly the same level of robustness. However, the band gap and spectrum no longer maintain symmetry, and the influence of the bulk states on the edge states becomes noticeable at a disorder strength smaller than that in the chiral case, Fig.~\ref{fig:fig1sm}.

\newpage
\section{Robustness for chains with odd number of sites in the chiral case \texorpdfstring{$t/f=-1$}{t/f=-1}}

\begin{figure}[h]
    \centering
    \includegraphics[scale=0.6]{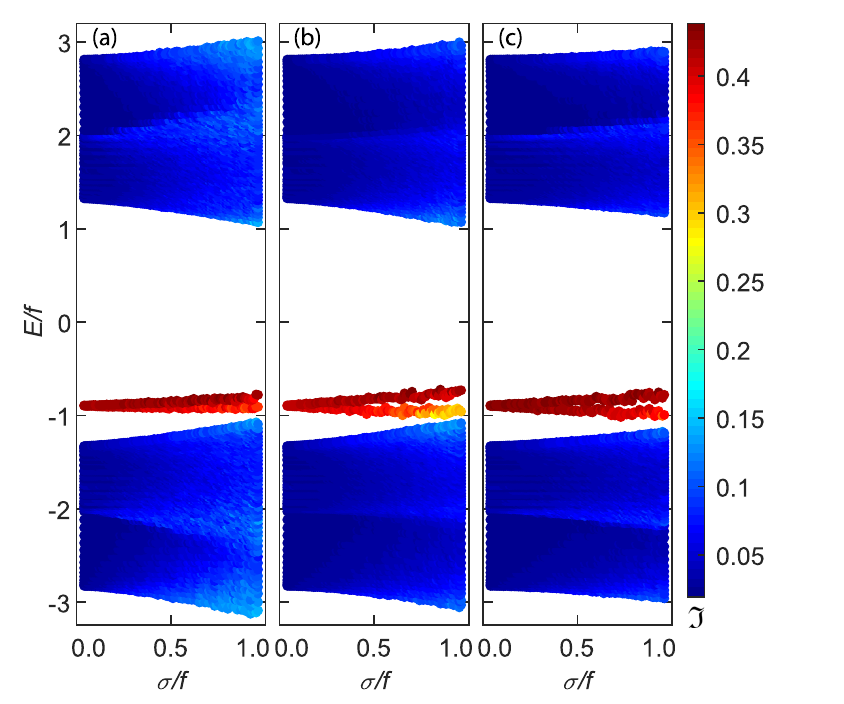}
    \caption{Energy eigenmodes computed for the zigzag system with $N=91$ sites, while setting~$v/f=1$ and $t/f=-1$. The three panels have been averaged over 30 random instances of a disorder $\sigma$ for the (a) next-neighbour (nn) interactions, (b) long-range (nnn) interactions, and (c) the on-site energy.}
    \label{fig:fig2sm}
\end{figure}

\noindent
In the figures presented in the main text, we illustrate the robustness properties of a chain consisting of an even number of sites in the chiral case. Interestingly, the edge states can be spotted also in chains with an odd number of sites, despite its degenerate energy levels. Higher disorders in this system result in a gradual spectral spreading of edge states that remains rather insignificant, e.g., less than $1\%$ of the band gap for $\sigma/f<0.2$. It increases by $4\%$ for $\sigma/f=0.5$, and reaches $10\%$ for $\sigma/f=1$, Fig.~\ref{fig:fig2sm}.

\clearpage
\newpage
\section{Robustness to losses as optimization task}
\label{sec:robustness_to_losses}

\begin{figure}[h]
  \centering
  \includegraphics[scale=0.57]{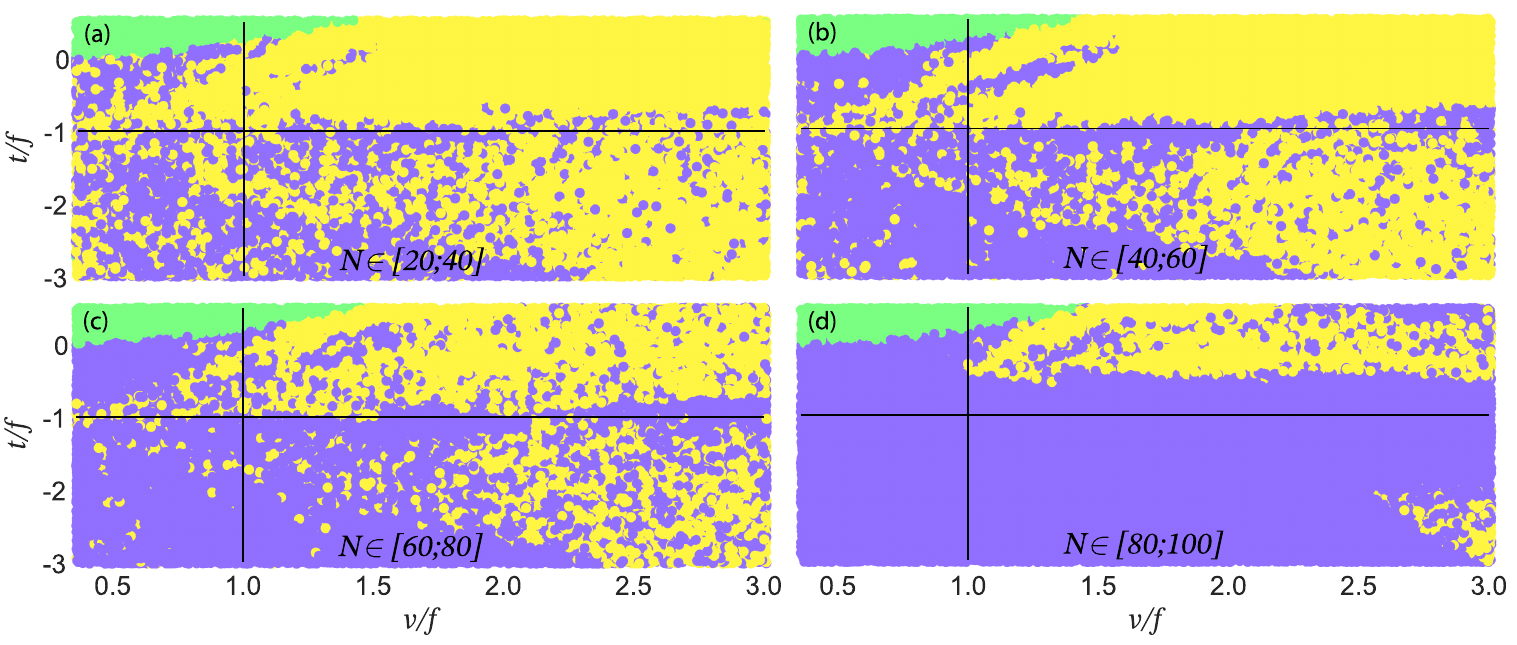}
  \caption{Robustness to losses for $v/f \in \left[0.3; 3.0 \right]$ and $t/f \in \left[ -3.0; 0.5\right]$. Green color represents the parameter region where the system does not support edge states, while yellow and violet regions correspond to areas with edge states, distinguishing between those that exhibit and those that do not exhibit robustness, respectively. For simplicity, the loss robustness is defined as the property of an edge state to remain selected for the same energy within the range $\gamma \in \left[0.0; 0.1\right]$. Each panel represents results based on approximately $5\times 10^4$ random points within the specified range of parameters $v/f$, $t/f$ and the system size (a) $N\in[20;40]$, (b) $N\in[40;60]$, (c) $N\in[60;80]$, and (d) $N\in[80;100]$. The horizontal solid line represents the chiral case $t/f=-1$, while the vertical one denotes the case of $v/f=1$.}
  \label{fig:stat}
\end{figure}

\noindent
To investigate the dependence of edge state robustness to losses in the range $\gamma \in [0.0, 0.1]$, we conducted calculations for $2.3\times 10^5$ parameter sets ${v/f, t/f, N}$, where $v/f \in [0.3, 3.0]$, $t/f \in [-3.0, 0.5]$, and $N \in [10, 100]$. Each set underwent an initial assessment for the presence of edge states using the IPR. The results are depicted in Fig.~\ref{fig:stat}, where the sets lacking edge states are marked in green.

Subsequently, the sets with edge states were analyzed for their ability to maintain well-localized edge states at the same energy while increasing $\gamma$ from $0$ to $0.1$. The ones exhibiting this property are labeled yellow, while those without it are marked with violet color. For clarity, all sets were grouped by their $N$ value. The black solid lines in Fig.~\ref{fig:stat} represent the parameter choices for the figures from the main text.

From these results, it is evident that the robustness of edge states to losses is directly correlated with $N$. The green areas in the panels of Fig.~\ref{fig:stat} align with the parabolic prediction from the topological invariant. Another observation is that most sets exhibit robustness within the ranges $t/f \in [-1.0, 0.5]$ and $v/f \in [1.0, 3.0]$. However, such strong near interactions lead to a reduction in the gap between the edge state and the closest energy band.

\clearpage

\section{Simultaneous presence of all types of disorder and losses in systems with positive and negative couplings}

\noindent
We would like to highlight that the studied system supports edge states when $t$ is both positive and negative, and these states remain robust in the presence of a certain level of disorders and losses. However, it is essential to consider the distinct nature of these edge states. Systems with different choices of $v/f$ and $t/f$ exhibit variations in localization and energy gap between the edge state and the closest bulk state. This variability is of significance for diverse implementations and applications.

\begin{figure}[h]\centering
  \includegraphics[scale=0.56]{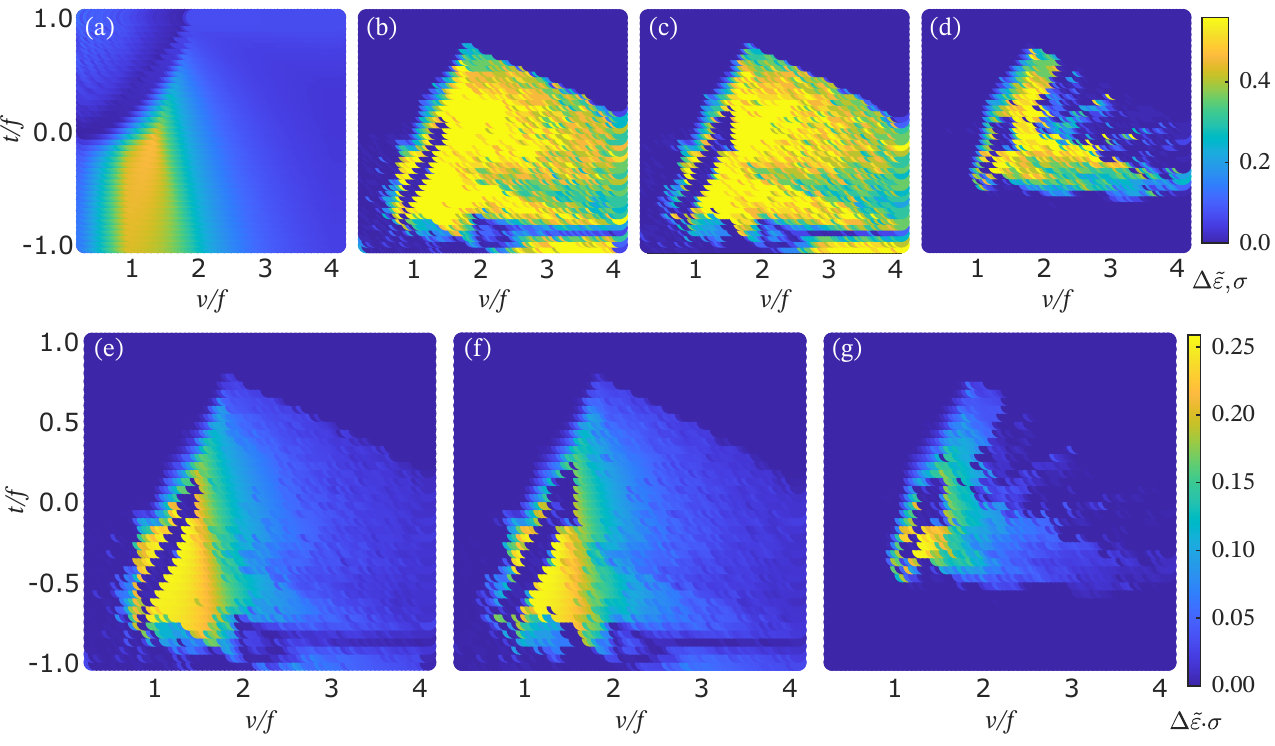}
  \caption{Optimization of the band gap size and robustness against simultaneous presence of all types of disorder and losses, with equal strength $\sigma$, computed for $v/f \in \left[0.3; 4.1 \right]$ and $t/f \in \left[ -1; 1\right]$. (a) Gap size $\varepsilon_{\text{edge state}}-\varepsilon_{\text{bulk state}}$ computed for the relative distance between the zigzag's edge state and the closest band, normalized to $\max(\lvert t/f\rvert, \lvert v/f\rvert,1)$. Panels (b)--(d) depict the maximally possible strength $\sigma$ of disorder and losses, while maintaining the edge states of the zigzag model at the same ($\pm15\%$) energy level and with an equivalent level of localization. The sizes of the systems in (b), (c), and (d) are $N=50$, $N=70$, and $N=90$, respectively. In (e)--(g), the results of element-wise multiplying the data from plot (a) by values from (b)--(d) are presented. Remarkably, the optimal choice for the most stable and remote edge states from the band edge can be achieved in models with negative couplings. However, for specific purposes, where the distance between the edge state and the band is less critical, it is possible to operate solely with positive couplings to attain a good level of robustness.}
  \label{fig:band}
\end{figure}

In Fig.~\ref{fig:band}a, the system displays the largest energy gap, $\Delta\tilde{\varepsilon}=(\varepsilon_{\text{edge state}}-\varepsilon_{\text{bulk state}})/\max(\lvert v/f\rvert,\lvert t/f\rvert,1)$, with normalization aimed at ensuring the largest couplings in the Hamiltonian are less than or equal to 1. Notably, a substantial portion of the brightest region resides in the diagram's negative $t/f$ quadrant.

Panels (b)--(d) illustrate the potential strength $\sigma$ when the system continues to support edge states at the same ($\pm15\%$) energy level and with identical localization. Disorder and losses are introduced to the system Hamiltonian as $H_{\sigma}=H+\sigma(D+i D')$ with operators $D$ and $D'$, respectively. Here, the reduction in the parameter region, where the system maintains robust edge states as its size increases, is primarily associated with a decrease in robustness to losses, as predicted in  Section~\ref{sec:robustness_to_losses}.

Figs.~\ref{fig:band}e--g represent the elementwise multiplication of panel (a) with panels (b)--(d), respectively. These figures distinctly indicate that the best edge states (considering localization, distance to the closest band, and robustness) are observed in systems with negative couplings. They also reveal a possibility of having edge states in zigzag chains exclusively with positive couplings, albeit with certain limitations that require attention. This paves the way for further research and simplifies the potential manufacturing process of the samples.

\clearpage
\section{Multi-orbital waveguides implementation}

A fourfold symmetric waveguide can help overcome the degeneracy of the $d$ orbital mode, which typically occurs in circular waveguides. This feature significantly simplifies the design process for structures where near and long interactions between waveguide modes play an equal role in the formation of topological phases. Additionally, a fourfold symmetric waveguide can support multiple optical modes when the side size $D$ of the waveguide exceeds the critical single-mode condition. We have fixed the side size at $D=12.4$ $\mu m$, with the refractive index of the cladding and two kinds of waveguides set to $n_{cl}=1.54$, $n_s=1.542$, and $n_d=1.5468$, respectively. The incident wavelength is $\lambda = 1.55$ $\mu m$, ensuring the same propagation constant $\beta_0=6.2464 \mu m^{-1}$ for both modes $s$ and $d$.

The main design issue is optimizing the coupling strength, which forms the interaction between the upper and lower layers of zigzag structures. The upper layer consists only of $s$ mode waveguides, while the lower layer comprises both $s$ and $d$ mode waveguides, as illustrated in Fig.~7(b) of the main manuscript. Fixing the distances between the centers of waveguides at $d_x=20$ $\mu m$ and $d_y=16$ $\mu m$ reduces the optimization problem to selecting the optimal shift parameter $\Delta$ from Fig.~7(d).

The coupling amplitudes between the decoupled modes of the fourfold symmetric waveguides are obtained by integrating the individual modes~\citeSM{Liu:17}:

\begin{equation}
    c_{mn}=  \pi \epsilon_0 c_0 \iint \sqrt{n^2_m-n^2_{cl}} \textbf{E}^*_m \cdot \textbf{E}_n \sqrt{n^2_n-n^2_{cl}}dx dy
\end{equation}

Here, $\textbf{E}_n$ represents the n-th normalized field of eigenstate of the isolated waveguides, and $n_n$ is the refractive index of the n-th isolated waveguide. With this, the coupling strength between waveguides of the upper layer is $c_1=2.26\cross 10^{-4}\mu m^{-1}$, and for the lower layer, it is $c_2=-2.88\cross 10^{-4}\mu m^{-1}$. The optimal choice for the shift value is $D=4 \mu m$, resulting in $c_2=1.73 \cross 10^{-4}\mu m^{-1}$ and $c_3=1.71 \cross 10^{-4}\mu m^{-1}$.

These geometric parameters allow for a match between FEM simulations and coupling mode theory analyses with normalized couplings $f=c_1/|c_4|=0.785$, $v=c_2/|c_4|\approx c_2/|c_4|=0.6$, $t=c_4/|c_4|=-1$. Theoretical models support robustness against all types of disorder.

\begin{figure}[h]
  \centering
  \includegraphics[scale=0.75]{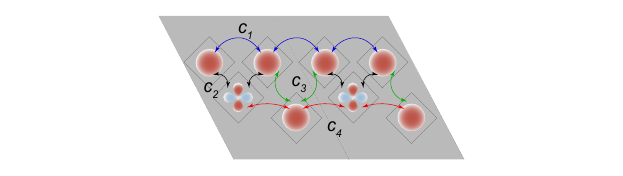}
  \caption{Design and couplings of the multi-mode waveguide structure. Blue and red indicate couplings $c_1$ and $c_4$ between waveguides in the upper and lower layers, respectively. Black and green represent couplings $c_2$ and $c_3$ between waveguides from different layers, where $c_2$ is between waveguides with $s$ and $d$ modes, and $c_3$ is between waveguides with $s$ modes.}
  \label{fig:structure}
\end{figure}

\bibliographystyleSM{unsrt}
\bibliographySM{our1}

\end{document}